\documentclass[aps,prx,twocolumn,superscriptaddress,longbibliography]{revtex4-1}
\usepackage{graphicx}% Include figure files
\usepackage{amsmath}
\usepackage{amssymb}
\usepackage[T1]{fontenc}
\usepackage[dvipsnames]{xcolor}
\begin{document}
\draft

\title{Prospects of forming high-spin polar molecules from ultracold atoms}

\author{Matthew D. Frye}
\affiliation{Joint Quantum Centre (JQC) Durham-Newcastle, Department of
Chemistry, Durham University, South Road, Durham, DH1 3LE, United Kingdom.}
\author{Simon L. Cornish}
\affiliation{Joint Quantum Centre (JQC) Durham-Newcastle, Department of
Physics, Durham University, South Road, Durham, DH1 3LE, United Kingdom.}
\author{Jeremy M. Hutson}
\email{j.m.hutson@durham.ac.uk} \affiliation{Joint Quantum Centre (JQC)
Durham-Newcastle, Department of Chemistry, Durham University, South Road,
Durham, DH1 3LE, United Kingdom.}

\date{\today}

\begin{abstract}
We have investigated Feshbach resonances in collisions of high-spin atoms such as Er and Dy with
closed-shell atoms such as Sr and Yb, using coupled-channel scattering and bound-state
calculations. We consider both low-anisotropy and high-anisotropy limits. In both regimes we find
many resonances with a wide variety of widths. The wider resonances are suitable for tuning
interatomic interactions, while some of the narrower resonances are highly suitable for
magnetoassociation to form high-spin molecules. These molecules might be transferred to short-range
states, where they would have large magnetic moments and electric dipole moments that can be
induced with very low electric fields. The results offer the opportunity to study
mixed quantum gases where one species is dipolar and the other is not, and open up important
prospects for a new field of ultracold high-spin polar molecules.
\end{abstract}

\maketitle

%\section{Introduction}

Magnetic Feshbach resonances play a crucial role in the study of ultracold atoms
\cite{Chin:RMP:2010}. They can be used to control interaction strengths by varying the scattering
length \cite{Moerdijk:1995}. This tunability has opened up many applications in the
many-body physics of strongly interacting systems \cite{Bloch:many:2008}. Tunable Feshbach
resonances also form the basis of magnetoassociation \cite{Kohler:RMP:2006}, and have been crucial
in the formation of ultracold ground-state molecules composed of alkali-metal atoms
\cite{Ni:KRb:2008, Takekoshi:RbCs:2014, Molony:RbCs:2014, Wang:2015, Park:NaK:2015, Guo:NaRb:2016}.
These molecules are now opening up new areas of research into dipolar physics \cite{Gorshkov:2011,
Yan:2013, Hazzard:2014, Seesselberg:PRL:2018}, quantum simulation and computation
\cite{DeMille:2002, Ni:Swap:2018, Blackmore:2019, Sawant:qudit:2019}, and controlled chemistry
\cite{Balakrishnan:FH2:2001, Tscherbul:JCP:2006, Hu:2019}.

There is now much interest in creating ultracold molecules from different species and with
different properties. In particular, molecules with both electric and magnetic dipole moments will
open up new possibilities for designing quantum many-body systems \cite{Micheli:2006, Baranov:2012}
and for tests of fundamental symmetries \cite{Safronova:2018}. In this context, there has been
significant work towards producing ultracold $^2\Sigma$ molecules, both by direct laser cooling
\cite{McCarron:2018, Anderegg:2018, Collopy:2018, Caldwell:2019} and by association of alkali-metal
and closed-shell atoms \cite{Zuchowski:RbSr:2010, Muenchow:2011, Brue:AlkYb:2013,
Guttridge:2p:2018, Barbe:RbSr:2018, Yang:CsYb:2019}. However, the former method still produces only
relatively low densities and is highly system-specific, while the latter method is experimentally
challenging due to the sparse and very narrow Feshbach resonances in such systems.

In this paper we investigate Feshbach resonances in collisions of lanthanide atoms with large
orbital angular momentum $l$, such as Er ($^3$H$_6$) or Dy ($^5$I$_8$), with closed-shell ($^1$S)
atoms, such as Yb or Sr. All four of these elements can be cooled to ultracold temperatures
\cite{Takasu:2003, Fukuhara:2009, Stellmer:2009, Mingwu:2011, Mingwu:2012, Aikawa:2012,
Aikawa:2014} and have numerous abundant isotopes including both bosons and fermions. The large
masses result in high densities of molecular states near threshold. These states
produce many resonances with a wide variety of widths. Some of them will be suitable for forming
molecules by magnetoassociation, and others will be suitable for tuning interactions in atomic
mixtures.

The variety of isotopes available should allow the formation of both bosonic and fermionic
molecules. These molecules will inherit the large magnetic moments of Er and Dy It should be
possible to transfer them to short-range states using Stimulated Raman Adiabatic Passage (STIRAP),
as has been achieved for alkali-metal dimers \cite{Ni:KRb:2008, Takekoshi:RbCs:2014,
Molony:RbCs:2014, Wang:2015, Park:NaK:2015, Guo:NaRb:2016}. In these short-range states they are
likely to have substantial electric and magnetic dipole moments. The magnetic dipole
interactions will be far larger than for currently available ultracold molecules, and systems built
from the molecules may exhibit new types of phase transition resulting from the competition between
electric and magnetic interactions.

The short-range states will be characterized by the projection $\Omega$ of the electronic (orbital
and spin) angular momentum along the internuclear axis. For molecules in states with $|\Omega|>0$,
the existence of small $\Omega$-type doubling splittings will allow the electric dipole moment to
be oriented with small electric fields \cite{Bethlem:IRPC:2003}. This is a particular benefit for
the high-spin molecules considered here, since $\Omega$-doubling splittings decrease fast as
$|\Omega|$ increases. Such molecules are analogous to symmetric tops; they will allow simulation of
important problems in quantum magnetism \cite{Wall:magnets:2013, Wall:2015} and may also open up
new opportunities for scalable quantum computers \cite{Yu:2019}. These advantages are not offered
by molecules formed from pairs of atoms in S states, even if they have high spin
\cite{Zaremba:2018}, since there is no $\Omega$-doubling for $\Sigma$ states.

The Feshbach resonances will also allow the study of quantum-degenerate Bose-Bose, Bose-Fermi and
Fermi-Fermi mixtures with tunable interactions. These mixtures will have the new feature that one
component interacts by dipole-dipole interactions, while the other does not. The mixtures will
exhibit rich phase behavior, including superfluidity and supersolidity
\cite{Capogrosso-Sansone:2011} and bound dipolar droplets \cite{Luis:2013}. They will open up new
possibilities for the study of imbalanced Fermi gases, which may exhibit exotic phase behavior and
allow the exploration of new pairing mechanisms \cite{Gubbels:2013, Baarsma:2016,
Ravensbergen:2018}.

\section{Theoretical methods and interaction potentials}\label{sec:theor}

The collision Hamiltonian for any pair of atoms $A$ and $X$ may be written
\begin{equation}
\frac{\hbar^2}{2\mu}\left(-\frac{d^2}{dR^2}+\frac{\hat{L}^2}{R^2}\right)+\hat{h}_A+\hat{h}_X+\hat{V}(R),
\end{equation}
where $R$ is the interatomic distance, $\mu$ is the reduced mass, and $\hat{L}$
is the angular momentum operator for the relative motion. $\hat{h}_A$ and
$\hat{h}_X$ are the Hamiltonians of the separated atoms \footnote{For
both operators and quantum numbers, we follow the common collisional convention
of using lower-case letters for individual atoms and upper-case letters for
the interacting pair.}, including external
fields if necessary, and $\hat{V}(R)$ is the interaction operator.

In the present work we consider the closed-shell atom $A$ to be structureless,
so set $\hat{h}_A=0$. The states of heavy open-shell atoms such as the
lanthanides Er and Dy are generally best represented using $j$-$j$ coupling
rather than Russell-Saunders coupling, but here we are interested principally
in the lowest spin-orbit component of the ground state, with the maximum total
angular momentum $j$ allowed by the orbital configuration. Other atomic states
(even excited spin-orbit states) are too high in energy to produce resonances
and play little role. We therefore choose to represent the atomic states with
quantum numbers $l$ and $s$ for the total orbital and spin angular momenta,
with corresponding operators $\hat{l}$ and $\hat{s}$, and the atomic
Hamiltonian as
\begin{equation}
\hat{h}_X=a^\textrm{so}_X\hat{l}\cdot\hat{s}+g_\textrm{l}(\hat{l}_z
+g_\textrm{s}\hat{s}_z)\mu_\textrm{B}B,
\end{equation}
where $B$ is the magnetic field oriented along the $z$ axis and $g_s$ and $g_l$ are the electron
spin and orbital g-factors, both defined to be positive. We take
$a^\textrm{so}_\textrm{Dy}=-516.779\times hc$ cm$^{-1}$ and
$a^\textrm{so}_\textrm{Er}=-1159.7215\times hc$ cm$^{-1}$, which reproduce the splitting of the
lowest two spin-orbit states \cite{Martin:1978}. In general there can also be hyperfine terms in
the single-atom Hamiltonians, but for simplicity, we focus on isotopes without nuclear spin in this
paper; we choose $^{164}$Dy, $^{174}$Yb, $^{166}$Er, and $^{88}$Sr, which are the most abundant
isotopes.

The interaction operator $\hat{V}(R)$ is a function of both orbital and spin coordinates. It can in
principle depend on external field, but any such dependence is neglected here. We choose to write
it in the form
\begin{equation}
\hat{V}(R) = \hat{V}_\textrm{space}(R) + \hat{V}_\textrm{spin}(R),
\end{equation}
where $\hat{V}_\textrm{space}(R)$ contains all terms independent of the spin degrees of freedom.
Because of cylindrical symmetry, $\hat{V}_\textrm{space}(R)$ is diagonal in $\lambda$, the
projection of $l$ onto the internuclear axis. We use the resolution of the identity $1 =
\sum_{l,\lambda} | l \lambda \rangle \langle l \lambda |$. Atomic states with $l\ne6$
for Dy or $l\ne5$ for Er are too high in energy to contribute, so we include only a single value of
$l$ in each case and the expansion reduces to
\begin{equation}
\hat{V}_\textrm{space}(R) = \sum_\lambda |l \lambda \rangle \langle l \lambda | V_\lambda(R),
\end{equation}
where $\lambda$ takes values from $-l$ to $l$ and $V_\lambda(R)$ is independent of the sign of
$\lambda$. The differences between the potential curves $V_\lambda(R)$ may be viewed as anisotropy
in $\hat{V}_\textrm{space}(R)$. The operator $\hat{V}_\textrm{spin}(R)$ might include
terms such as the dependence of $a^\textrm{so}_X$ on $R$; for simplicity we neglect it here,
although it may be needed when making quantitative comparisons between experiment and theory at a
later stage.

Very little is known about the interactions in the systems we consider here.
However, we can make reasonable estimates on physical grounds. The attractive
parts of the potentials are likely to be dominated by dispersion forces, with
little contribution from chemical bonding, because the outermost s orbitals are
filled and the partially filled f shells of the lanthanides are submerged. At
long range the interaction potentials are of the form $V(R)=-C_6/R^6$. In
physical terms, differences between the potentials $V_\lambda(R)$ may come from
differences in either the attractive or the repulsive part. To model these
separately, we represent the interaction potentials in Lennard-Jones form,
\begin{align}
V_\lambda(R) &= C_{12,\lambda} R^{-12} - C_{6,\lambda} R^{-6}  \nonumber\\
&=D_{\textrm{e},\lambda} \left [ \left( \frac{R_{\textrm{e},\lambda}}{R} \right)^{12} - 2
\left( \frac{R_{\textrm{e},\lambda}}{R} \right)^6 \right],
\end{align}
where $C_{12,\lambda}$ is the repulsive coefficient and $D_{\textrm{e},\lambda}$ is the well depth
at equilibrium distance $R_{\textrm{e},\lambda}$. More sophisticated forms of $V_\lambda(R)$ could be used,
but would not change the qualitative conclusions.

We obtain parameters for Er+Sr and Dy+Yb from combination rules based on the
interaction potentials for the corresponding homonuclear systems. Values of the
$C_6$ coefficients for all the homonuclear systems have been obtained
experimentally \cite{Stein:2010, Borkowski:2017} or theoretically
\cite{Porsev:2014, Safronova:2012, Lepers:Er:2014, Li:Dy:2016}. We obtain
values of the isotropic coefficients $C_6^{(0)}$ for Er+Sr and Dy+Yb from
Tang's combination rule \cite{Tang:1969}, for which we use dispersion
coefficients from \cite{Stein:2010, Borkowski:2017, Lepers:Er:2014, Li:Dy:2016}
and the atomic polarizabilities from \cite{Schwerdtfeger:2019}, giving 2092 and
2359 $E_\textrm{h} a_0^6$, for Dy+Yb and Er+Sr respectively.

Er and Dy have tensor and vector polarizabilities that result in anisotropic $C_6$ coefficients.
The ratio of the anisotropic coefficient $C_6^{(2)}$ to the isotropic one is comparable to the
ratio of the tensor and scalar static polarizabilities of the open-shell atom; values for this
range from 0.016 \cite{Chu:Lanth:2007} to 0.018 \cite{Lepers:Er:2014} for Er and from 0.005
\cite{Li:Dy:2016} to 0.026 \cite{Chu:Lanth:2007} for Dy. We base our main calculations on a ratio
$C_6^{(2)}/C_6^{(0)}=0.017$ for both systems, but have explored the effects of variations.

The interaction potential for Sr+Sr has been studied in detail spectroscopically; the well depth
and equilibrium distance are known precisely to be 1081.64 cm$^{-1}$ and 4.672 \AA, respectively,
and the potential supports 63 vibrational levels for $^{88}$Sr \cite{Stein:2008, Stein:2010}. For
Yb+Yb there is considerable variation in the well depth between different levels of theory
\cite{Tecmer:2019}, but the dependence of the near-threshold bound states on isotopic mass shows
that it supports 72 vibrational levels \cite{Kitagawa:2008, Borkowski:2017}. We base our
calculations on the depth 739.73 cm$^{-1}$ obtained by Borkowski et al.\ \cite{Borkowski:2017}.
Petrov et al.\ \cite{Petrov:2012} carried out electronic structure calculations of one component of
the potential for Dy+Dy and obtained a well depth of 785.7 cm$^{-1}$. To obtain interaction
potentials for Er+Sr and Dy+Yb, we estimate the isotropic well depths $D_\textrm{e}$ as the
geometric mean of those for the two homonuclear systems. For this we estimate the well depth for
Er+Er as 471.1 cm$^{-1}$, obtained from the well depth for Dy+Dy \cite{Petrov:2012} scaled by the
square of the ratio of the $C_6$ coefficients. The resulting well depths for the heteronuclear
systems are about 760 cm$^{-1}$ for Dy+Yb and 710 cm$^{-1}$ for Er+Sr.

To model the effects of long-range anisotropies, we choose a set of
coefficients $C_{6,\lambda}$ that correspond to the required value of
$C_6^{(2)}$,
\begin{equation}
C_{6,\lambda} = C_6^{(0)} + C_6^{(2)} g_2(l,\lambda),
\end{equation}
where
\begin{equation}
g_k(l,\lambda) = (-1)^\lambda
(2l+1)
\left(\begin{matrix}l & k & l \cr 0 & 0 & 0\end{matrix}\right)
\left(\begin{matrix}l & k & l \cr -\lambda & 0 & \lambda\end{matrix}\right).
\label{eq:gaunt}
\end{equation}
We obtain an isotropic repulsive coefficient from
\begin{equation}
C_{12}^{(0)}=(C_6^{(0)})^2/4D_\textrm{e}.
\end{equation}
We fix $C_{12,\lambda}$ at $C_{12}^{(0)}$ for all $\lambda$, so that
$D_{\textrm{e},\lambda}$ is different for each $\lambda$.

Short-range anisotropies are less well understood. They might be much stronger,
because of the effects of higher-order dispersion, chemical bonding or
repulsive forces. To model this, we choose a set of well depths
$D_{\textrm{e},\lambda}$ and fix $C_6$ at its isotropic value, so that
$C_{12,\lambda}$ is different for each $\lambda$. The isotropic potential
curves described above support 52 (64) vibrational levels for Dy+Yb (Er+Sr).
Because of this, a 3.1\% (3.8\%) scaling of the potential is sufficient to
shift the scattering length through a complete cycle
\cite{Hutson:theory-cold-colls:2009}. It thus seems likely that the scattering
lengths for different values of $\lambda$ are essentially random. Varying
$D_\textrm{e}$ with $C_6$ unchanged is not a simple scaling, but we
nevertheless find that a scaling of $D_\textrm{e}$ by 4.7\% (5.6\%) produces a
full cycle in the scattering length. Our strategy is thus to choose values of
$D_{\textrm{e},\lambda}$ randomly from a range from 0.93 to 1.07 (0.91 to 1.09)
times the starting value. This range is chosen so that even the isotropic
potential (essentially the mean $V_\lambda(R)$) is random. This produces
anisotropies of higher order than the dispersion anisotropy above, with $k>2$
in Eq.\ \eqref{eq:gaunt}. It should be emphasized that this probably represents
the \emph{maximum} coupling that is likely to arise from short-range
anisotropy, and the coupling may turn out to be weaker.

We perform scattering calculations using the \textsc{molscat} package \cite{mbf-github:2020,
molscat:2019} and bound-state calculations with the \textsc{bound} package \cite{mbf-github:2020,
bound+field:2019}. Both packages use coupled-channel methods, in which the total wavefunction is
expanded in a basis set for all coordinates except the interatomic distance $R$. This gives a set
of coupled differential equations with respect to $R$. The methods are similar to those of Refs.\
\cite{Gonzalez-Martinez:LiYb:2013, Gonzalez-Martinez:2015} but without terms that correspond to
$\hat{h}_A$. We use a partially coupled basis set $|(ls)jm_j\rangle |L M_L\rangle$, where $m_j$ and
$M_L$ are the projections of $j$ and $L$ onto the axis of the magnetic field (the $z$ axis). Matrix
elements of the interaction potential are calculated as in Ref.\ \cite{Gonzalez-Martinez:H+F:2013}.
The only conserved quantum numbers are the projection of the total angular momentum onto the $z$
axis, $M_\textrm{tot}=m_j+M_L$, and the total parity $(-1)^{l+L}$.
Since the atomic spin-orbit splittings for both
Dy and Er are substantially larger than the potential well depths, we restrict our basis functions
to those for the lowest spin-orbit state, with $j=l+s$. This restriction means that the spin-orbit
coupling in $\hat{h}_X$ has no effect except to shift the whole system by a constant energy. We
include all functions with $L$ up to $L_\textrm{max}$. For the case of long-range anisotropy, the
basis set converges quickly with respect to $L$ and we choose $L_\textrm{max}=8$. For short-range
anisotropy, which can be stronger, we include basis functions up to $L_\textrm{max}=2j$. This is
because, in a representation where $j$ and $L$ are coupled to form a total angular momentum $J$,
the s-wave channel ($L=0$) at the lowest threshold corresponds to $J=j$; representing all states of
this total $J$ requires functions up to $L=2j$. In the resulting calculations, the widest
resonances are well converged. For strong anisotropy, adding higher-$L$ basis functions shifts some
of the narrower resonances and adds new narrow ones.

The coupled equations are solved using log-derivative propagators. For scattering calculations, the
log-derivative matrix is propagated outwards from $R_\textrm{min}=2.5$~\AA\ to
$R_\textrm{mid}=20$~\AA\ using the diabatic log-derivative method of Manolopoulos
\cite{Manolopoulos:1986, Manolopoulos:1993:Johnson} with a fixed step size of 0.005~\AA, then from
$R_\textrm{mid}$ to $R_\textrm{max}=2000$~\AA\ using the log-derivative Airy propagator of
Alexander and Manolopoulos with a variable step size \cite{Alexander:1984, Alexander:1987}. The
solution is matched to asymptotic boundary conditions at $R_\textrm{max}$ to obtain the scattering
matrix $\textbf{S}$. The scattering length is calculated as $a(k)=(ik)^{-1}(1-S_{00})/(1+S_{00})$,
where $k=\sqrt{2\mu E}/\hbar$ is the wavevector and $S_{00}$ is the diagonal S-matrix element in
the incoming s-wave channel for the chosen value of $m_j$. The kinetic energy $E$ in the incoming
channel is set to 100 nK. For bound-state calculations, one log-derivative matrix is propagated
outwards from $R_\textrm{min}$ to $R_\textrm{mid}$, and another is propagated inwards from
$R_\textrm{max}$ to $R_\textrm{mid}$. \textsc{bound} finds eigenenergies by searching for the
energies at which these solutions can be matched at $R_\textrm{mid}$ as described in
\cite{Hutson:CPC:1994}.

\begin{figure*}[tbh]
\includegraphics[width=0.99\linewidth]{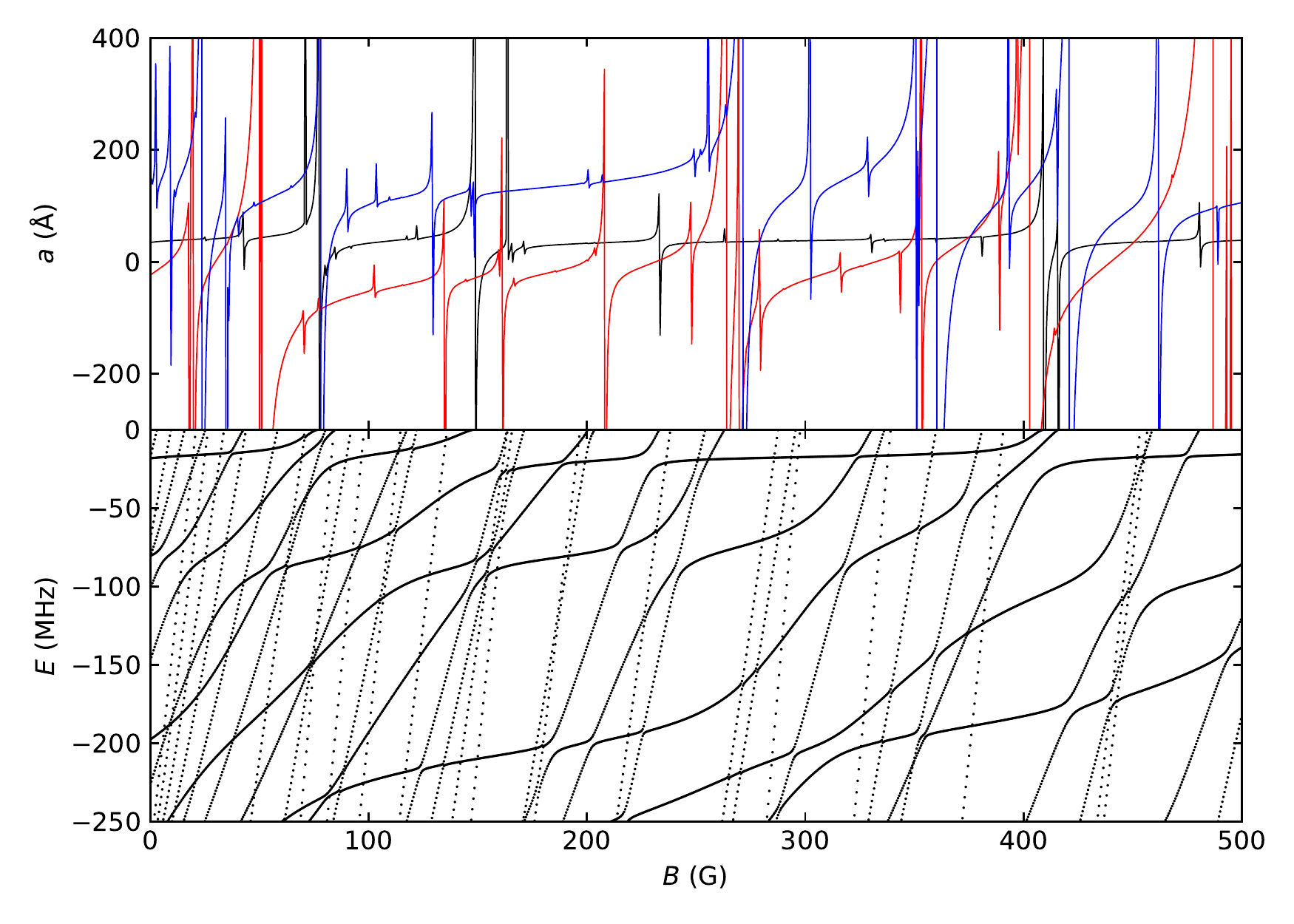}
\caption{\label{fig:DyYb_att} Top: Scattering lengths as a function of magnetic
field for Dy+Yb with long-range anisotropy $C_6^{(2)}/C_6^{(0)} = 0.017$ for
three different values of the scattering length $a_\textrm{iso}=54$~\AA\
(black), $-80$~\AA\ (red), and 168~\AA\ (blue). Bottom: The corresponding
near-threshold bound states for $a_\textrm{iso}=54$~\AA.}
\end{figure*}

\section{Feshbach resonances and bound states}

A magnetic Feshbach resonance occurs when a bound state is tuned across a scattering threshold with
a magnetic field $B$. Molecules may be formed by sweeping the field across a resonance; this is
referred as magnetoassociation. Resonances may also be used to control interactions by tuning the
scattering length.

For an isolated resonance at the lowest threshold, the scattering length $a(B)$ has a
characteristic pole at the resonance position $B_\textrm{res}$ \cite{Moerdijk:1995},
\begin{equation}
\label{eq:a_res}
a(B)=a_\textrm{bg}\left(1-\frac{\Delta}{B-B_\textrm{res}}\right),
\end{equation}
where $a_\textrm{bg}$ is the background scattering length and $\Delta$ is the resonance width.

We have carried out coupled-channel calculations to characterize Feshbach resonances at the lowest
threshold of Dy+Yb ($^5$I$_8$, $m_j=-8$). We first consider interaction potentials with a purely
long-range anisotropy characterized by $C_6^{(2)}/C_6^{(0)}=0.017$. The details of the results
depend on the scattering length $a_\textrm{iso}$ of the isotropic potential, which is at present
unknown. We have therefore performed calculations for 100 different interaction potentials, with
values of $C_{12}^{(0)}$ that vary over a range of $\sim 4.7\%$,  covering the full range of
$a_\textrm{iso}$ but with
similar well depths. The upper panel of Figure \ref{fig:DyYb_att} shows the s-wave scattering
length for three representative potentials with $a_\textrm{iso}=54$~\AA, $-80$~\AA\ and 168~\AA.

We have characterized all the Feshbach resonances that exist below 500~G for each interaction
potential, using the method of Frye and Hutson \cite{Frye:resonance:2017, Frye:quasibound:2020}. There are typically 50
resonances for each potential, and we obtain values of $B_\textrm{res}$, $\Delta$ and
$a_\textrm{bg}$ for each resonance. It should be noted that the
resonance width $\Delta$ can be artificially large if $a_\textrm{bg}$ is particularly small. A
better measure of the strength of the resonant pole, which retains dimensions of magnetic field, is
the normalized width \cite{Yang:CsYb:2019}
\begin{equation}
\bar{\Delta} = a_\textrm{bg} \Delta / \bar{a},
\label{eq:delta-bar}
\end{equation}
where $\bar{a}=(2\mu C_6/\hbar^2)^{1/4}\times 0.4779888\dots$ is the mean scattering length of
Gribakin and Flambaum \cite{Gribakin:1993} and is 40.3~\AA\ for $^{164}$Dy+$^{174}$Yb and 37.7~\AA\ for
$^{166}$Er+$^{88}$Sr.

All the interaction potentials produce many Feshbach resonances with a wide distribution of widths.
The widest resonances are due to bound states with predominantly $L=2$ character, and the narrower
ones to states with predominantly $L=4$ character. However, as seen below, there is substantial
mixing between states of different $L$. Very narrow resonances with $L>4$ exist, and some of them
are just visible in Figure \ref{fig:DyYb_att}. In Supplemental Material \cite{Supp_DyYb} we plot the number of
resonances with $\bar{\Delta} > \bar{\Delta}_\textrm{min}$ as a function of
$\bar{\Delta}_\textrm{min}$: 95\% of the interaction potentials with $C_6^{(2)}/C_6^{(0)}=0.017$ have at
least 10 resonances with $\bar{\Delta} > 0.1$~G at fields below 500~G.

The details of the resonances depend on $a_\textrm{iso}$, but choosing a different value of
$C_{12}^{(0)}$ that produces a different number of bound states but the same $a_\textrm{iso}$
produces very similar results. Most resonances shift by less than 0.2
G, and their widths change by less than 20\%.

The lower panel of Figure \ref{fig:DyYb_att} shows the near-threshold bound states responsible for
the resonances in the upper panel for the case of $a_\textrm{iso}=54$~\AA. The near-horizontal line
near $E=-20$ MHz is due to a state that is predominantly $L=0$, $m_j=-8$ in character. This has
broad avoided crossings with $L=2$ states and much narrower ones with $L=4$ states; the slopes of
these states increase and the strengths of the crossings decrease as $m_j$ increases from $-8$.
More generally there are strong crossings between pairs of states with $\Delta L \le 2$ and $\Delta
m_j \le 2$ and weaker crossings due to higher-order interactions otherwise. For Dy+Yb the
characteristic energy $\bar{E} = \hbar^2/(2\mu\bar{a}^2) = h\times 3.69$ MHz. There is a bound
state with $L=0$ (2) in the top $36.1\bar{E}$ ($93.4\bar{E}$) of the well for any value of
$a_\textrm{iso}$ \cite{Gao:2000}, so in the top 133 (344) MHz here. The $g$-factor for the
$^5$I$_8$ state is $g_j = 1.25$, so the relative gradient of states with $m_j=-8$ and $-6$ is
$2g_j\mu_\textrm{B}/h = 3.50$ MHz/G. The lowest-field $L=2$ resonance is expected to occur at a
field below $93.4 \bar{E}/2g_j\mu_\textrm{B} = 98$ G for any value of $a_\textrm{iso}$.

We have repeated the calculations for a smaller long-range anisotropy $C_6^{(2)}/C_6^{(0)}=0.0085$,
which is half the value in Fig.\ \ref{fig:DyYb_att}. Representative examples of
scattering lengths and bound-state energies are shown in Supplemental Material \cite{Supp_DyYb}.
The bound-state diagram shows considerably less mixing between different states. The resonance
positions change very little for the same value of $a_\textrm{iso}$, but the widths scale
approximately as $|C_6^{(2)}|^2$ for resonances due to $L=2$ states and as $|C_6^{(2)}|^4$ for
those due to $L=4$ states. This indicates that a long-range anisotropy of this magnitude may be
viewed as operating perturbatively. In this case all 100 potentials considered had at
least 3 resonances with $\bar{\Delta} > 0.1$~G below 500~G. The long-range
anisotropies considered here probably represent approximately the \emph{minimum} degree of coupling
that is likely in these systems.

We have carried out analogous calculations on Er+Sr. The results are qualitatively similar, and
representative examples are shown in the Supplemental Material \cite{Supp_DyYb}. In
this case, $\bar{E}=h\times 6.18$~MHz and $g_j = 1.17$, so the lowest-field $L=2$ resonance is
expected to occur below $93.4 \bar{E}/2g_j\mu_\textrm{B} = 176$ G for any value of
$a_\textrm{iso}$. The resulting densities of bound states and resonances are somewhat lower than
for Dy+Yb.

\begin{figure*}[tbh]
\includegraphics[width=0.99\linewidth]{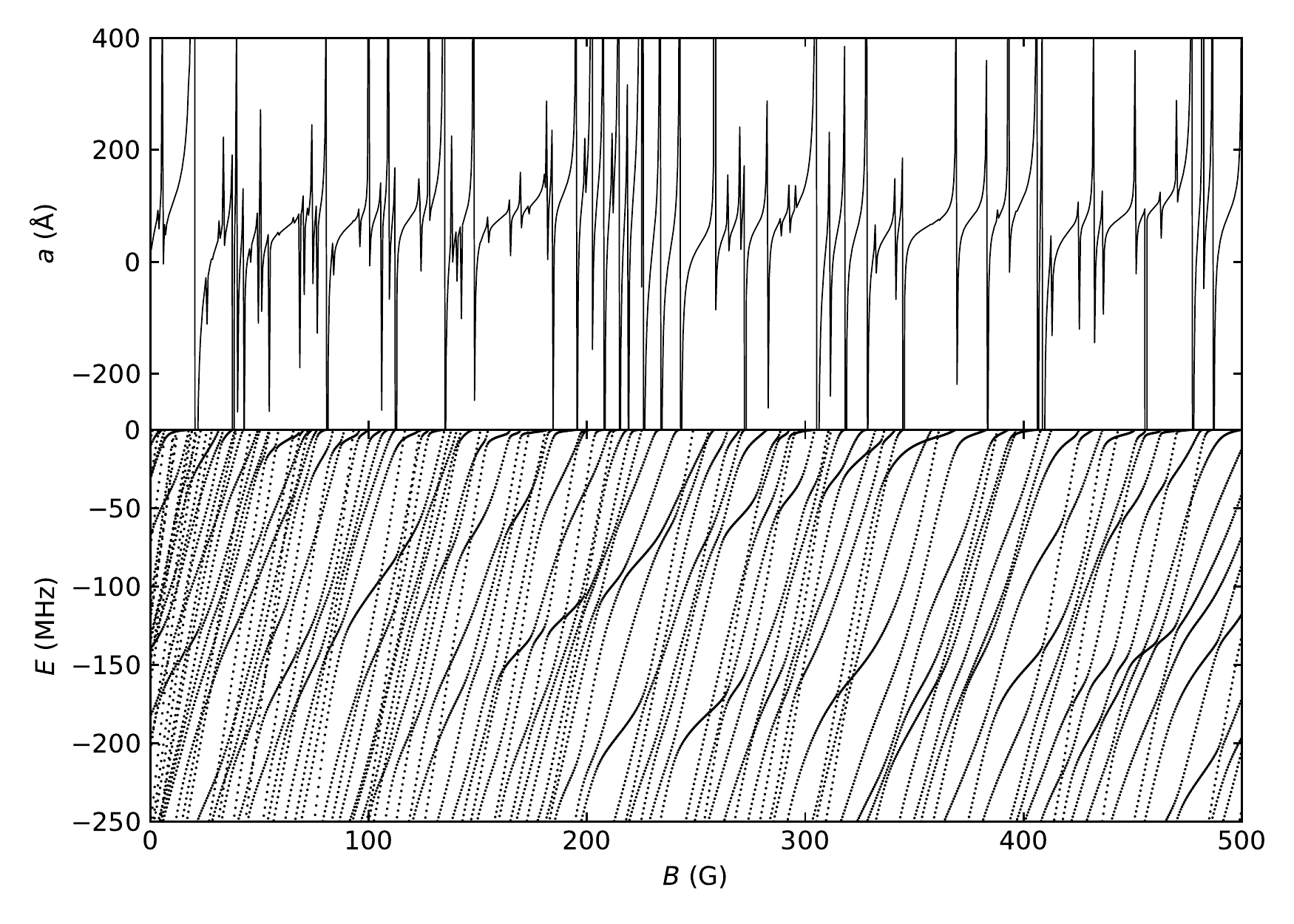}
\caption{\label{fig:DyYb_rep} Top: Scattering lengths as a function of field
for Dy+Yb with short-range anisotropy as described in the text. Bottom: The
corresponding near-threshold bound states.}
\end{figure*}

We next consider interaction potentials with short-range anisotropies, chosen as
described in section \ref{sec:theor}. The upper panel of Figure \ref{fig:DyYb_rep} shows the
scattering length for Dy+Yb ($m_j=-8$) for a representative example, and the lower panel shows the
near-threshold bound states. Further examples are shown in Supplemental Material \cite{Supp_DyYb}. These
interaction potentials are much more strongly anisotropic than those based on long-range anisotropy
above, and their higher-order anisotropies cause direct couplings with larger values of $\Delta L$
and $\Delta m_j$. They all produce many Feshbach resonances with a wide variety of widths. $L$ is
now very poorly conserved, so the resonances are not limited to those for $L<6$. Nevertheless, it
may be seen in the lower panel of Fig.\ \ref{fig:DyYb_rep} that there are well-defined states that
maintain their character through many avoided crossings. This demonstrates that the energy-level
pattern has structure even in this high-anisotropy regime.

We have characterized all the resonances that exist below 500~G for 100 different
interaction potentials of this type. In Supplemental Material \cite{Supp_DyYb} we plot the number of resonances with
$\bar{\Delta} > \bar{\Delta}_\textrm{min}$ as a function of $\bar{\Delta}_\textrm{min}$. We find
that 95\% of interaction potentials with short-range anisotropy have at least 60 resonances with
$\bar{\Delta} > 0.1$~G at fields below 500~G.

One advantage of these systems is the existence of many different isotopic combinations. The
reduced mass may be adjusted over a range of 4.7\% for Dy+Yb or 4.6\% for Er+Sr. This changes the
patterns of resonances and bound states in significantly different ways in the different regimes of
anisotropy. For weak anisotropy a change in reduced mass $\mu$ is almost the same as scaling the
isotropic potential; it changes the isotropic scattering length $a_\textrm{iso}$ in a smooth way.
Changing the mass of the open-shell (closed-shell) atom by 2 changes $\mu$ by 0.6\% (0.6\%) for
Dy+Yb or 0.4\% (1.5\%) for Er+Sr; this is sufficient to change $a_\textrm{iso}$ considerably, but
it is also possible to change the isotope of both atoms to achieve a much smaller change in reduced
mass. For strong anisotropy, however, the situation is different; in this case the positions of the
widest resonances shift relatively smoothly, but even a change of 0.1\% in $\mu$ causes substantial
changes in the relative positions of narrower resonances. Measurements on different isotopic pairs
will thus provide a powerful way to distinguish between the different regimes of anisotropy.

\begin{figure*}[tbh]
\includegraphics[width=0.99\linewidth]{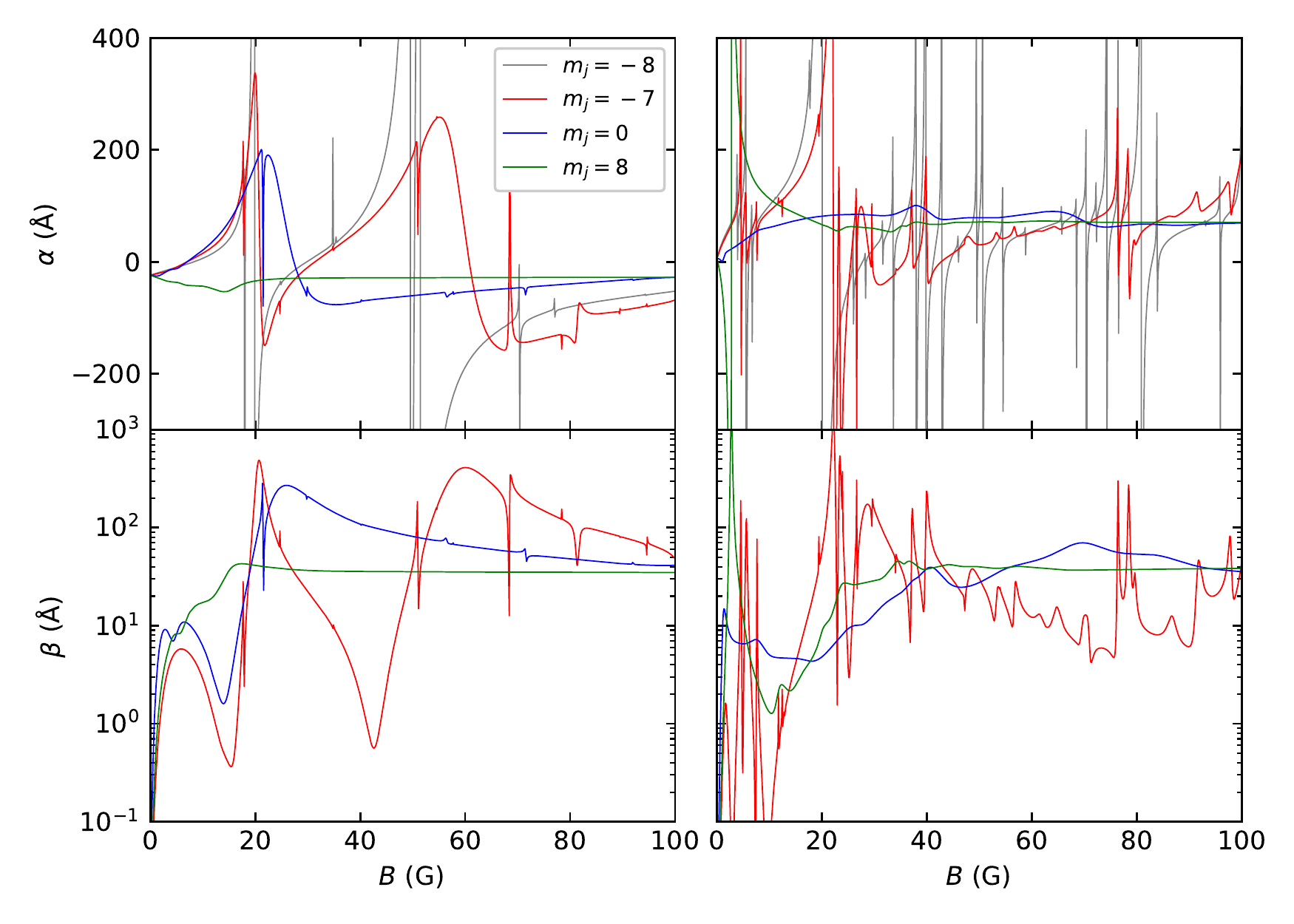}
\caption{\label{fig:DyYb_decay} Complex scattering lengths $a=\alpha-i\beta$ as
a function of magnetic field for Dy+Yb with Dy in excited Zeeman states
$m_j=-7$ (red), 0 (blue), and 8 (green); $m_j=-8$ (grey) results are also shown
for comparison. Left-hand panels: Interaction potential with long-range
anisotropy $C_6^{(2)}/C_6^{(0)} = 0.017$ for $a_\textrm{iso}=-80$~\AA. Right:
Interaction potential with short-range anisotropy as described in the text.}
\end{figure*}

We have also calculated scattering lengths for collisions of open-shell atoms
in excited Zeeman states, $m_j>-j$. Examples for Dy+Yb with both long-range and
short-range anisotropies are shown in Fig.\ \ref{fig:DyYb_decay}. Additional
examples are given in Supplemental Material \cite{Supp_DyYb}. The scattering
length is complex, $a=\alpha-i\beta$, because of inelastic decay to lower
channels. The corresponding rate coefficient for 2-body inelastic
decay in s-wave scattering is $k_2^\textrm{inel} \approx 2h\beta/\mu$. The
resonances show oscillations in both $\alpha$ and $\beta$
\cite{Hutson:res:2007, Frye:resonance:2017}, rather than poles as at the lowest
threshold. For long-range anisotropies, the resonances are relatively weakly
decayed for $m_j=-7$, with oscillations in $\alpha$ and peaks in $\beta$ of
amplitude up to 500 \AA. The amplitudes of the oscillations decrease with
increasing $m_j$, as the number of open channels increases. For short-range
anisotropies the decay is substantially stronger; the resonances are
progressively washed out as $m_j$ increases, and $\beta$ approaches the value
of $\bar{a}$ expected in the ``universal" limit of 100\% loss at short range
\cite{Idziaszek:PRL:2010, Frye:2015}. In this case there are visible resonances
in $\alpha$ and $\beta$ at low field even for the uppermost channel with
$m_j=j$; these arise from quasibound states trapped behind centrifugal barriers
for channels with $m_j<j$ and $L>0$ \cite{Kato:ybresonances:2013}.

\section{Extension to Bose-Fermi and Fermi-Fermi mixtures}

Our calculations have dealt with systems where both nuclei are spin-zero bosons and the
corresponding atoms are bosonic. Fermionic isotopes also exist, with non-zero nuclear spin $i$, and
Bose-Fermi and Fermi-Fermi mixtures are of great interest. The fermionic isotopes include
$^{161}$Dy ($i=5/2$), $^{163}$Dy ($i=5/2$), $^{167}$Er ($i=7/2$), $^{87}$Sr ($i=9/2$), $^{171}$Yb
($i=1/2$), $^{173}$Yb ($i=3/2$). These elements have no stable bosonic isotopes with $i>0$.

Including nuclear spin introduces substantial additional complexity. Each additional spin is an
extra source of angular momentum, and there are additional terms in the Hamiltonian from nuclear
Zeeman effects and from couplings between the additional spin and the existing angular momenta $j$
and $L$. The cases with a nuclear spin on the lanthanide or on the closed-shell atom are quite
different.

A nuclear spin on the closed-shell atom is uncoupled to $j$ and $L$ in the separated atoms and only
weakly coupled in the molecular states that cause Feshbach resonances. A moderate magnetic field is
sufficient to decouple it even for the molecule, with a well-defined projection quantum number
$m_i$ onto the field axis. Under these circumstances the nuclear spin acts as a spectator; the
pattern of resonances and widths is essentially the same as without the spin, with small shifts
dependent on $m_i$ due to nuclear Zeeman effects. There may be additional narrow resonances due to
couplings involving the nuclear spin, as exist for a closed-shell atom with nuclear spin
interacting with an alkali-metal atom \cite{Brue:LiYb:2012, Barbe:RbSr:2018, Yang:CsYb:2019}.

A nuclear spin on the lanthanide atom is strongly coupled to the atomic $j$ even in the separated
atoms, with hyperfine splittings up to a few GHz \cite{Childs:Dy:1970, Childs:Er:1983}. There are
multiple zero-field atomic states that interact and mix in a magnetic field. The states for a
single $j$ may be viewed as a more complicated version of the states of an alkali-metal atom in a
magnetic field. For a high-spin atom in an S state, such as Cr or Eu \cite{Tomza:2013, Tomza:2014},
interaction with the closed-shell atom produces only a single electronic state. The resulting
atomic and molecular states would then be coupled only by the $R$-dependence of the hyperfine
coupling constants; this is relatively weak, and would produce only narrow resonances, as are found
for an alkali-metal atom interacting with a closed-shell atom \cite{Zuchowski:RbSr:2010,
Brue:AlkYb:2013, Barbe:RbSr:2018, Yang:CsYb:2019}. In the present case, however, the multiple
electronic states resulting from the anisotropy of the lanthanide atom play the same role as the
singlet and triplet states in alkali pairs \cite{Chin:RMP:2010}. The combined effect of hyperfine
coupling, Zeeman mixing and multiple electronic states will produce complex patterns of Feshbach
resonances, over and above those due to the fine structure and anisotropy alone. Characterizing
these resonances in detail is beyond the scope of the present work, but they are likely to produce
additional opportunities for tuning mixtures of fermionic lanthanide atoms and closed-shell atoms
and producing molecules by magnetoassociation.

When there are nuclear spins on both atoms, as is the case for Fermi-Fermi mixtures of lanthanide
atoms and closed-shell atoms, both the effects described above will come into play, producing a
rich and complex pattern of Feshbach resonances. Nevertheless, the calculations described in the
present paper are likely to represent a lower bound for the number of resonances that will exist.

\section{Experimental Possibilities}

All four atomic species considered here have been cooled to quantum degeneracy
for both bosonic and fermionic isotopes. From a laser-cooling perspective, the
highly magnetic lanthanide atoms are remarkably similar to their closed-shell
cousins, despite their complex atomic structure. Indeed, the procedure for
creating Er Bose-Einstein condensates \cite{Aikawa:2012} was inspired by the
work on Yb \cite{Takasu:2003,Fukuhara:2007}. They also have similar
polarizabilities in the wavelength range commonly used for optical traps (1030
to 1070~nm), so that co-trapping and simultaneous evaporative cooling of
co-trapped samples should be straightforward. The atoms are thus highly compatible
for experiments on mixtures.

The behavior of the mixtures will be complicated by the dense and chaotic
intraspecies resonance structure of Er and Dy themselves \cite{Frisch:2014,
Baumann:2014, Maier:ChaosErDy:2015}. However, the positions of these resonances
are well documented and the majority have widths below 100\,mG. Two possible
approaches may be envisaged. Since the closed-shell atoms are unaffected by the
intraspecies resonances of Dy or Er, and have no intraspecies resonances
themselves, they may be used to probe the interspecies resonances, even in
regions where intraspecies resonances exist. Alternatively, if the closed-shell
atom is in large excess, three-body losses due to the intraspecies resonances
may be sufficiently suppressed to work with the interspecies resonances. In any
case there are numerous windows (up to $\sim$$1$~G wide) that are free from
intraspecies resonances and suitable for working with interspecies resonances.

The resonance-free windows should also be ideal for evaporative cooling of both
species to quantum degeneracy. The recent successful creation of
dual-degenerate gases of Dy and Er \cite{Trautmann:2018} is extremely
encouraging in this respect. Simultaneous evaporative cooling requires a
favorable interspecies scattering length, which may be achieved either by
tuning across a wide resonance with magnetic field or (in the low-anisotropy
case) by a suitable choice of isotopes to achieve the required background
scattering length.

Once degeneracy is achieved, it should be possible to load the mixture into a
3D optical lattice. Such a lattice provides a very sensitive environment for
the detection of Feshbach resonances \cite{Mark:2018} and an ideal platform for
molecule formation. A lattice with only one atom of Dy or Er per site
eliminates the complication of intraspecies resonances and allows
magnetoassociation at any desired magnetic field.

Magnetoassociation is usually most effective at relatively narrow resonances,
with widths around 100~mG to 1~G. Many such resonances exist in both limiting
regimes of anisotropy. In the low-anisotropy case they are mostly due to bound states with
$L=4$ or 6. The background scattering length for these resonances is moderately
tunable with magnetic field, as seen in Fig.\ \ref{fig:DyYb_att}, but is
centered around the isotropic scattering length $a_\textrm{iso}$. If necessary,
$a_\textrm{iso}$ may be selected by choosing an appropriate isotopic
combination. In the high-anisotropy case there are narrow resonances suitable
for magnetoassociation, with a wide variety of background scattering lengths
even for a single interaction potential (or isotopic combination). In this
case, careful selection of isotopes may be unnecessary.

Transferring molecules formed by magnetoassociation to short-range (low-lying) states with electric
dipole moments will be a challenge. Nevertheless, it can probably be achieved by STIRAP, as has
been possible for a variety of alkali-metal dimers. Detailed spectroscopy of the intermediate
electronic states available for STIRAP will be required, but is outside the scope of the present
paper.

The short-range states of these molecules will probably be best described by
Hund's case (c) \cite{Veseth:1973}. Here the atomic total angular momentum $j$
is projected onto the internuclear axis with projection $\Omega$, which can
take values from $-j$ to $j$. For each value of $\Omega$, the total angular
momentum $J$ can take values $J\ge|\Omega|$ in integer steps. If $j$ is a good
quantum number, states $\Omega$ and $\Omega'$ with $\Omega-\Omega'=\pm1$ are
coupled by Coriolis matrix elements proportional to
$[(j(j+1)-\Omega\Omega')(J(J+1)-\Omega\Omega')]^{1/2}$, which ultimately
uncouple $j$ from the internuclear axis at high $J$.

In a magnetic field, each state will be split into $2J+1$ components with
space-fixed projection $M$ from $-J$ to $J$. The g-factor for Hund's case (c)
is \cite{McDonald:2018}
\begin{equation}
\frac{\Omega^2}{2J(J+1)}\left(g_s+g_l+(g_s-g_l)\frac{s(s+1)-l(l+1)}{j(j+1)}\right).
\end{equation}
The quantity in parentheses is approximately 7/3 for $^3$H$_6$ and 5/2 for $^5$I$_8$.
For $J=|\Omega|$ there will be
$2|\Omega|+1$ such states, equally spaced at low field. It should be
straightforward to transfer population between these states and create coherent
superpositions of them with microwave pulses, as has been achieved for
alkali-metal dimers \cite{Gregory:RbCs-microwave:2016}. The large number of
states available offer the opportunity to create fully controllable
high-dimensional quantum systems, which may be used as qudits for quantum
computation \cite{Sawant:qudit:2019}.

In zero electric field, states with positive and negative values of $\Omega$ combine to form pairs
of states of opposite parity. These parity eigenstates have no permanent dipole moment. However,
they are separated only by coupling to $\Omega=0$, which exists for only one parity combination.
They are thus nearly degenerate, and become more closely so with increasing $|\Omega|$. For states
with high $|\Omega|$, it requires only a very small electric field to mix the parity eigenstates
($\Omega$-doublets) and form oriented states with dipole moment $\mu_v \Omega M/[J(J+1)]$. Here
$\mu_v$ is the body-fixed molecular dipole moment and $\Omega$ and $M$ are signed quantities; the
space-fixed dipole depends on their relative sign. These states are analogous to
those of symmetric top molecules, which are expected to have important applications in the
simulation of quantum magnetism \cite{Wall:magnets:2013, Wall:2015} and in quantum computing
\cite{Yu:2019}.

The interactions of high-spin atoms with closed-shell atoms are substantially simpler
than those with alkali-metal atoms or with themselves. As described above, the interaction of
Dy($^5$I$_8$) with a closed-shell atom is characterized by at most 7 potential curves,
$V_\lambda(R)$. If the short-range anisotropy is weak, the scattering lengths for the 7 curves may
be related to one another; even if not, it is likely to be feasible to extract this many scattering
lengths from experimental measurements on Feshbach resonances and near-threshold bound states.
Interactions of high-spin atoms with alkali-metal atoms \cite{Gonzalez-Martinez:2015} require twice
as many potential curves, because there are two possible spin multiplicities for each $\lambda$.
Interactions of high-spin atoms with themselves require many more potential curves
\cite{Petrov:2012}.

\section{Conclusions}

We have considered Feshbach resonances in collisions of an open-shell high-spin
atom (Dy or Er) with a closed-shell atom (Yb or Sr). We have developed model
interaction potentials for Dy+Yb and Er+Sr, and calculated scattering lengths
and the positions of near-threshold bound states as a function of magnetic
field. For both systems we have found numerous Feshbach resonances, with a
variety of widths, at moderate magnetic fields.

The couplings responsible for Feshbach resonances depend on the anisotropy of
the interaction potential. We have considered two limiting regimes of
anisotropy. In the first limit, we consider anisotropy due entirely to
dispersion forces, which arises from the tensor polarizability of the
open-shell atom. This produces quite weak anisotropies in the long-range
potential for atoms like Dy and Er, in which the unpaired f electrons lie
mostly inside the outermost s electrons. The resulting interaction produces
direct couplings with selection rule $\Delta L\le 2$, where $L$ is the
end-over-end angular momentum of the colliding pair. In this regime the
strongest resonances in s-wave scattering are due to bound states with $L=2$;
we show that at least one such resonance must occur below 98~G for Dy+Yb and
below 176~G for Er+Sr. Additional broad resonances occur at somewhat higher
fields, and there are also narrower resonances due to bound states with $L>4$.
In the second limit, we consider much stronger anisotropy that may exist due to
higher-order dispersion forces, chemical bonding or repulsive anisotropy. In
this regime there is much stronger coupling between different values of $L$,
such that $L$ is no longer even nearly conserved. This produces many more
Feshbach resonances, but still with a wide variety of widths.

The two regimes of anisotropy that we have considered span the range of likely physical behavior
for these systems. The long-range model represents the minimum coupling that is likely to exist,
while the short-range model represents the maximum. For both models we predict many Feshbach
resonances at experimentally accessible magnetic fields, with a variety of widths.
Additional resonances may occur when one or both atoms have nuclear spin.

The resonances predicted here have a wide range of possible applications. The wider resonances are
very suitable for tuning interspecies interactions, both to achieve properties desirable for
formation of dual degenerate gases and to investigate the novel properties of mixtures of dipolar
and non-dipolar species \cite{Capogrosso-Sansone:2011, Luis:2013, Baarsma:2016}.
Bose-Bose, Bose-Fermi and Fermi-Fermi mixtures are accessible. Further tuning may be achieved by
selecting from the wide variety of isotopic combinations available. The narrower resonances may be
used for magnetoassociation to form high-spin molecules that inherit the large magnetic moments of
Er and Dy. These may in turn be transferred to short-range states where they will still have large
magnetic moments. In addition, there will be near-degenerate pairs of $\Omega$-doublets, of
opposite parity, which will be easily mixed with very small electric fields to form oriented states
with significant space-fixed electric dipole moments.

The general properties of the resonances predicted here will hold for a variety
of systems. They apply to any combination of a heavy atom with non-zero spin
and high orbital angular momentum and a heavy closed-shell atom. The densities
of near-threshold states and of the resulting resonances will be lower if
either atom is significantly lighter, but the general considerations still
apply for atoms of mass 20 or more. They apply to transition-metal atoms as
well as lanthanides or actinides. They do not apply to open-shell atoms in S
states or to very light atoms such as He and first-row elements.

In summary, we believe that this study reveals new directions for the
study of strongly interacting quantum gases and opens up a new field of high-spin
dipolar molecules. We have demonstrated the feasibility of producing such
molecules from laser-cooled atoms, and outlined some of their properties and
potential applications.

\begin{acknowledgments}
We are grateful to Alex Guttridge for valuable discussions. This work was
supported by the U.K. Engineering and Physical Sciences Research Council
(EPSRC) Grants No.\ EP/N007085/1, EP/P008275/1 and EP/P01058X/1.
\end{acknowledgments}

\bigskip\noindent
\emph{Note added:} After completion of this work, we learnt of parallel work on
the related system Er+Yb that also predicts many resonances at low field
\cite{kosicki:ErYb:2020}.

\bibliography{../all,Dy_Yb,Dy_Yb-SLC}

\end{document}